# Damping Effect of Electromagnetic Radiation and Time-Dependent Schrödinger Equation

Ji Luo

*Department of Photoelectric Technology,*
*Chengdu University of Information Technology, Chengdu 610225, Sichuan, China*
*Surface Physics Laboratory, Department of Physics,*
*Fudan University, Shanghai 200433, China*

The inexactness of the time-dependent Schrödinger equation of a charged particle in an external electromagnetic field is discussed in terms of the damping effect of the radiation. A possible improvement is to add a nonlinear term representing this effect to the linear Schrödinger equation. Conditions for the nonlinear term are investigated and it is demonstrated that the obtained nonlinear Schrödinger equation may present state evolutions similar to the wave-function reduction and transitions between stationary states.

PACS numbers: 03.65.Ta, 02.30.Tb, 03.65.Ge

The standard time-dependent linear Schrödinger equation (TDLSE) determines the wave-function evolution of a quantum system in such a way that a superposition of the eigen-states of the Hamiltonian remains a superposition. Nevertheless, the actually observed physical realities are the stationary eigen-states, the eigen-energies, and the transitions between the eigen-states. Although the TDLSE can present resonant transitions for the perturbation field with an appropriate frequency, it fails to present spontaneous transitions [1]. In general, an arbitrary initial state does not necessarily evolve into an eigen-state. The underlying physics is beyond the TDLSE, and one must resort to the so-called wave-function reduction, which is regarded as a process different from the state evolution governed by the TDLSE [2-4].

Existing important improvements resort to the macroscopic apparatus and the environment [5-8]. Nevertheless, the system itself is also crucial because the TDLSE may be inexact. Suppose a particle with charge $q$ and mass $m$ moves in an external electromagnetic field $\vec{E}$, $\vec{B}$ with the scalar potential $\varphi(\vec{r},t)$ and the vector potential $\vec{A}(\vec{r},t)$ satisfying $\vec{E} = -\nabla\varphi - \partial\vec{A}/\partial t$, $\vec{B} = \nabla \times \vec{A}$.

The standard TDLSE of the particle is

$$i\hbar(\partial \psi / \partial t) = \hat{H}\psi, \quad (1)$$

where $\psi(\vec{r},t)$ is the wave function and $\hat{H}$ is the Hamiltonian operator

$$\hat{H} = (1/2m)(i\hbar\nabla + q\vec{A})^2 + q\varphi. \quad (2)$$

$\hat{H}$ is obtained according to its counterpart of the non-relativistic classical particle $H = (1/2m)(\vec{P} - q\vec{A})^2 + q\varphi$ [1,9]. The way is to substitute the canonical momentum $\vec{P} = m\vec{v} + q\vec{A}$ with the operator $-i\hbar\nabla$, where $\vec{v}$ is the velocity. Further backward, the classical Hamiltonian is derived form the Lagrangian $L(\vec{r},\vec{v}) = mv^2/2 - q(\varphi - \vec{v}\cdot\vec{A})$. Finally, the Lagrangian has this form because its Lagrange equation $(d/dt)(\partial L/\partial v_w) - \partial L/\partial w = 0$ with $w = x, y, z$ is equivalent to the Newton equation $m(d\vec{v}/dt) = q(\vec{E} + \vec{v}\times\vec{B})$.

A charged classical particle moving in acceleration radiates electromagnetic waves. Due to the radiation, the particle is subjected to an extra recoil force in addition to the Lorentz force $q(\vec{E} + \vec{v}\times\vec{B})$ [9]. However, the Newton equation that procedurally leads to the Hamiltonian operator of the quantum particle does not contain this recoil force. Consequently, TDLSE (1) is inexact in that it does not include the damping effect of the radiation.

Adding a nonlinear term representing the damping effect of the radiation to the TDLSE becomes a physically founded improvement. In general one has the following time-dependent nonlinear Schrödinger equation (TDNLSE)

$$i\hbar(\partial\psi/\partial t) = \hat{H}\psi + \hat{R}\psi, \quad (3)$$

where $\hat{R}$ is a Hermitian operator. As an improvement, TDNLSE (3) should retain important properties of TDLSE (1), must not contradict existing physical facts, should provide new results, and should shed light on some unsolved problems. We shall examine the nonlinear term according to these principles. We also employ the correspondence between the quantum particle and the classical one, that is, we investigate the nonlinear term in terms of its effect as the particle gradually becomes classical and the wave function reduces to a localized wave packet.

Firstly, $\hat{R}$ should satisfy

$$\psi(\hat{R}\psi)^* = \psi^*\hat{R}\psi. \quad (4)$$

According to this condition one obtains from TDNLSE (3) the same continuity equation as that from TDLSE (1), that is,

$$\partial\rho/\partial t + \nabla\cdot\vec{J} = 0, \quad (5)$$

where $\rho(\vec{r},t) = \psi\psi^*$ is the probability density and $\vec{J}(\vec{r},t) = -(i\hbar/2m)(\psi^*\nabla\psi - \psi\nabla\psi^*) - (q/m)\psi\psi^*\vec{A}$ is the usual current density [1]. This is because in calculating $\partial\rho/\partial t = \psi^*\partial\psi/\partial t + \psi\partial\psi^*/\partial t$ from TDNLSE (3), the nonlinear term disappears. Hence like TDLSE (1), TDNLSE (3) retains the normalization of the wave function for the normalized initial state, since Eq. (5) leads to $\partial\left(\int_\infty \rho dV\right)/\partial t = 0$. We hereafter consider normalized wave functions.

Secondly, the velocity operator of the particle is $\hat{\vec{v}} = -(i\hbar/m)\nabla - (q/m)\vec{A}$. Hence $\vec{J} = \text{Re}(\psi^*\hat{\vec{v}}\psi)$ and the average velocity $\langle\vec{v}\rangle = \int_\infty \psi^*\hat{\vec{v}}\psi dV = \int_\infty \vec{J}dV$. For a quasi-classical particle in a wave-packet state, it is the time derivative of the velocity that closely relates to the radiation effect. Hence the time derivatives $\partial\rho/\partial t$ and $\partial\vec{J}/\partial t$ play an important role in the nonlinear term.

Thirdly, for a time-independent external field, if the particle stays in a stationary state $\psi(\vec{r},t) = \phi(\vec{r})\exp[i\theta(t)]$ where $\theta(t)$ is a real function, both $\rho$ and $\vec{J}$ are independent of time and the particle has no radiations. This is the well established conclusion that the charged system in a stationary state does not emit electromagnetic waves. As an actually observed state, the stationary state must be a solution of TDNLSE (3). Since the stationary wave function satisfies TDLSE (1), $\hat{R}$ must satisfy

$$\hat{R}\psi = 0 \qquad (6)$$

for a stationary state. As a result, TDLSE (1) and TDNLSE (3) have the same stationary solutions $\phi_n(\vec{r})\exp(-iE_n t/\hbar)$ with $n = 0,1,2,\cdots$, where $\phi_n(\vec{r})$ are the orthonormal real eigen-functions of the time-independent $\hat{H}$ and $E_n$ are the corresponding eigen-values.

Finally, TDNLSE (3) should present a different wave-function evolution from that presented by TDLSE (1). For a time-independent $\hat{H}$, if the initial state is an eigen-state, the particle should remain in this eigen-state. If the initial state is a superposition of the eigen-states, the wave function should eventually evolve into a single eigen-state, since the observed state is always a stationary one. This conclusion should remain for a time-dependent field that tends to a time-independent one as time increases. Furthermore, for a superposed small perturbation field, TDNLSE (3) should present the spontaneous transition form the initial eigen-state to the final eigen-state, and for a superposed periodical field with an appropriate frequency, TDNLSE (3) should present the resonant transition between the corresponding eigen-states. In Ref. 10, the nonlinear term given by $\hat{R}\psi = \beta(\partial\rho/\partial t)\psi$ is investigated for $\vec{A} \equiv 0$, where $\beta$ is a constant. Numerical calculations demonstrate that this nonlinear term realizes the above expectations. We shall prove the expected conclusions generally for the appropriate nonlinear term and obtain more conditions for $\hat{R}$.

As the quantum particle gradually becomes classical, the nonlinear term in TDNLSE (3) gives the damping effect of a classical particle. For the classical particle, the more exact Newton equation is $m(d\vec{v}/dt) = \vec{F}_q + \vec{F}_r$, where $\vec{F}_q = q(\vec{E} + \vec{v} \times \vec{B})$ is the Lorentz force and $\vec{F}_r$ is the recoil force of the radiation. For the quantum particle, according to TDNLSE (3) the time derivative of $\langle \vec{v} \rangle$ is derived as

$$m(d\langle\vec{v}\rangle/dt) = \langle \vec{F}_q \rangle + \langle \vec{F}_r \rangle, \tag{7}$$

where

$$\langle \vec{F}_q \rangle = q\int_\infty \psi^*[\vec{E} + (1/2)(\hat{\vec{v}} \times \vec{B}) - (1/2)(\vec{B} \times \hat{\vec{v}})]\psi dV, \tag{8}$$

$$\langle \vec{F}_r \rangle = (m/i\hbar)\int_\infty \psi^*[\hat{\vec{v}}, \hat{R}]\psi dV, \tag{9}$$

and $[\hat{\vec{v}}, \hat{R}] = \hat{\vec{v}}\hat{R} - \hat{R}\hat{\vec{v}}$. This is the Ehrenfest theorem of TDNLSE (3) [11]. For a wave-packet state, Eq. (8) becomes $\langle \vec{F}_q \rangle = q\vec{E}\int_V \psi\psi^* dV + (q/2)\int_V \psi^*\hat{\vec{v}}\psi dV \times \vec{B} - (q/2)\vec{B} \times \int_V \psi^*\hat{\vec{v}}\psi dV$, where $\vec{E}$ and $\vec{B}$ is the field at the wave packet and $V$ is the volume of the wave packet. Hence $\langle \vec{F}_q \rangle$ reduces to the classical form $q(\vec{E} + \langle\vec{v}\rangle \times \vec{B})$, since $\int_V \psi\psi^* dV = 1$ and $\int_V \psi^*\hat{\vec{v}}\psi dV = \langle\vec{v}\rangle$. Accordingly, $\langle \vec{F}_r \rangle$ should be the classical recoil force of the radiation. For the classical particle with a moderate velocity, $|\vec{F}_r| \ll |\vec{F}_q|$ [9]. Hence one expects $|\langle \vec{F}_r \rangle| \ll |\langle \vec{F}_q \rangle|$.

Since the classical recoil force has never been exactly given [9], we consider the energy variation of both the classical particle and the quantum one. For the classical particle, we retain the usual energy definition $E = mv^2/2 + q\varphi$ even for the time-dependent field. Hence $dE/dt = d(mv^2/2)/dt + qd\varphi/dt$. According to the theorem of kinetic energy $d(mv^2/2)/dt = [q(\vec{E} + \vec{v} \times \vec{B}) + \vec{F}_r] \cdot \vec{v}$ and $d\varphi/dt = \nabla\varphi \cdot \vec{v} + \partial\varphi/\partial t$ one obtains

$$dE/dt = q[(\partial\varphi/\partial t) - (\partial\vec{A}/\partial t) \cdot \vec{v}] + \vec{F}_r \cdot \vec{v}, \tag{10}$$

where $-\vec{F}_r \cdot \vec{v}$ is the radiation power. For the quantum particle, the corresponding energy operator is the Hamiltonian operator $\hat{H} = (m/2)\hat{\vec{v}}^2 + q\varphi$ and the average energy is $\langle E \rangle = \int_\infty \psi^*\hat{H}\psi dV$. According to TDNLSE (3), the time derivative of $\langle E \rangle$ is derived as

$$d\langle E \rangle/dt = q\int_\infty [(\partial\varphi/\partial t)\rho - (\partial\vec{A}/\partial t) \cdot \vec{J}]dV + (1/i\hbar)\int_\infty \psi^*[\hat{H}, \hat{R}]\psi dV, \tag{11}$$

where $[\hat{H}, \hat{R}] = \hat{H}\hat{R} - \hat{R}\hat{H}$. For the wave-packet state, the first integral in Eq. (11) becomes $q(\partial\varphi/\partial t)\int_V \rho dV - q(\partial\vec{A}/\partial t) \cdot \int_V \vec{J}dV$, where $\partial\varphi/\partial t$ and $\partial\vec{A}/\partial t$ have values at the place of the wave packet and $V$ is the volume of the wave packet. Hence this integral reduces to the classical form $q[(\partial\varphi/\partial t) - (\partial\vec{A}/\partial t) \cdot \langle\vec{v}\rangle]$, since $\int_V \rho dV = 1$ and $\int_V \vec{J}dV = \langle\vec{v}\rangle$. Accordingly, the second integral in Eq. (11) should reduce to the negative of the radiation power of the classical particle. By a comparison of Eq. (11) and Eq. (10), one concludes that
$$P(t) = -(1/i\hbar)\int_\infty \psi^*[\hat{H},\hat{R}]\psi dV \quad (12)$$
is the radiation power of the quantum particle and for the wave-packet state,
$$P(t) = -\langle\vec{F}_r\rangle\cdot\langle\vec{v}\rangle. \quad (13)$$
Physically, the quantum particle does not radiate electromagnetic waves if and only if it stays in a stationary state. Hence $P(t)\equiv 0$ if and only if $\psi(\vec{r},t)=\phi_n(\vec{r})\exp(-iE_n t/\hbar)$ with $n=0,1,2,\cdots$, where $\phi_n(\vec{r})$ and $E_n$ are the eigen-functions and eigen-energies of $\hat{H}$ with $E_0\leq E_1\leq E_2\leq\cdots$.

Suppose the initial state $\psi(\vec{r},t_0)$ has the average energy $\langle E(t_0)\rangle$. According to Eqs. (11) and (12), one has
$$\langle E(t)\rangle - \langle E(t_0)\rangle = q\int_{t_0}^t\{\int_\infty[(\partial\varphi/\partial t)\rho - (\partial\vec{A}/\partial t)\cdot\vec{J}]dV\}dt - \int_{t_0}^t P(t)dt, \quad (14)$$
where $\int_{t_0}^t P(t)dt$ is the radiation energy. First we suppose the external field is time-independent. For a bound state, the wave function $\psi(\vec{r},t)$ can be expanded in terms of the eigen-functions $\phi_n(\vec{r})$ of the time-independent $\hat{H}$, that is, $\psi(\vec{r},t)=\sum_{n=0}^\infty C_n(t)\phi_n(\vec{r})$, where $C_n(t)$ are the coefficients satisfying $\sum_{n=0}^\infty |C_n(t)|^2=1$. Hence one has
$$\langle E(t)\rangle = \sum_{n=0}^\infty |C_n(t)|^2 E_n \geq E_0, \quad (15)$$
where $E_0$ is the ground-state energy. Since $\partial\varphi/\partial t=0$ and $\partial\vec{A}/\partial t=0$, according to Eqs. (14) and (15) one obtains
$$\int_{t_0}^t P(t)dt \leq \langle E(t_0)\rangle - E_0. \quad (16)$$
Hence $\int_{t_0}^t P(t)dt$ is bounded by an upper limit. This means that the particle cannot radiate electromagnetic waves endlessly and one has
$$\lim_{t\to+\infty} P(t)=0. \quad (17)$$
Because $P(t)\equiv 0$ holds only for stationary states, the particle eventually evolves into an eigen-state $\phi_k(\vec{r})\exp(-iE_k t/\hbar)$ with $k\in\{0,1,2,\cdots\}$. Finally, since $\partial\varphi/\partial t=0$ and $\partial\vec{A}/\partial t=0$, according to Eq. (14) the radiation energy is
$$\int_{t_0}^{+\infty} P(t)dt = \langle E(t_0)\rangle - E_k. \quad (18)$$
For the initial state that is a superposition of the eigen-states, this is the process similar to the wave-function reduction.

Next we suppose a time-dependent external field tends to a time-independent one as time increases. Then $\phi_n(\vec{r})$ and $E_n$ are respectively the eigen-functions and eigen-values of the final time-independent $\hat{H}$, and Eq. (15) holds asymptotically as $t\to+\infty$. Equations (14) and (15) leads to
$$\int_{t_0}^t P(t)dt \leq \langle E(t_0)\rangle - E_0 + q\int_{t_0}^t\{\int_\infty[(\partial\varphi/\partial t)\rho - (\partial\vec{A}/\partial t)\cdot\vec{J}]dV\}dt \quad (19)$$

for sufficient large $t$. As long as for all $t$,

$$\left|\int_{t_0}^{t}\{\int_\infty [(\partial\varphi/\partial t)\rho - (\partial\vec{A}/\partial t)\cdot\vec{J}]dV\}dt\right| \leq M, \tag{20}$$

where $M$ is a positive constant, $\int_{t_0}^{t} P(t)dt$ is bounded by an upper limit and Eq. (17) holds. Since $\partial\varphi/\partial t \to 0$ and $\partial\vec{A}/\partial t \to 0$ as $t \to +\infty$, condition (20) is satisfied [12]. The particle evolves into an eigen-state $\phi_k(\vec{r})\exp(-iE_k t/\hbar)$ with $k \in \{0,1,2,\cdots\}$. According to Eq. (14) the radiation energy is

$$\int_{t_0}^{+\infty} P(t)dt = \langle E(t_0)\rangle - E_k + q\int_{t_0}^{+\infty}\{\int_\infty [(\partial\varphi/\partial t)\rho - (\partial\vec{A}/\partial t)\cdot\vec{J}]dV\}dt. \tag{21}$$

Finally we suppose the external field is the superposition of a time-independent one and a time-dependent one. In this case $\phi_n(\vec{r})$ and $E_n$ are respectively the eigen-functions and eigen-values of $\hat{H}$ corresponding to the time-independent field. We consider two cases: First, the time-dependent field is a small perturbation that exists only for a limited period of time. This is a special case of what is discussed in the above paragraph, since $\partial\varphi/\partial t \to 0$, $\partial\vec{A}/\partial t \to 0$ as $t \to +\infty$. Hence condition (20) is satisfied and Eq. (17) holds. The particle evolves into an eigen-state $\phi_k(\vec{r})\exp(-iE_k t/\hbar)$ with $k \in \{0,1,2,\cdots\}$ and Eq. (21) gives the radiation energy. For the initial state $\phi_j(\vec{r})\exp(-iE_j t_i/\hbar)$ with $j \in \{0,1,2,\cdots\}$, according to Eq. (14) the radiation energy is

$$\int_{t_0}^{+\infty} P(t)dt = (E_j - E_k) + q\int_{t_0}^{+\infty}\{\int_\infty [(\partial\varphi/\partial t)\rho - (\partial\vec{A}/\partial t)\cdot\vec{J}]dV\}dt. \tag{22}$$

For a small perturbation field, the integral $\int_{t_0}^{+\infty}\{\int_\infty [(\partial\varphi/\partial t)\rho - (\partial\vec{A}/\partial t)\cdot\vec{J}]dV\}dt$ is small and one has $\int_{t_0}^{+\infty} P(t)dt \approx E_j - E_k$. This is the process similar to the spontaneous transition. Second, the time-dependent field is a periodic one with a large frequency. Numerical calculations demonstrate that for a small periodic field, the wave function evolves form the initial state $\phi_j(\vec{r})\exp(-iE_j t_i/\hbar)$ to a final one $\phi_k(\vec{r})\exp(-iE_k t/\hbar)$ with $k \in \{0,1,2,\cdots\}$ [10]. Furthermore, for the periodic field with the angular frequency $\omega = |E_k - E_j|/\hbar$, the wave function evolves in such a way that it alternates between the initial eigen-state $\phi_j(\vec{r})\exp(-iE_j t_i/\hbar)$ and the eigen-state $\phi_k(\vec{r})\exp(-iE_k t/\hbar)$ as time increases [10]. This is similar to the stimulated resonant transition between two eigen-states. The radiation energy is given by Eq. (22). Since $\partial\varphi/\partial t$ and $\partial\vec{A}/\partial t$ have a large frequency, one has $\int_{t_0}^{+\infty}\{\int_\infty [(\partial\varphi/\partial t)\rho - (\partial\vec{A}/\partial t)\cdot\vec{J}]dV\}dt \approx 0$ and $\int_{t_0}^{+\infty} P(t)dt \approx E_j - E_k$.

The exact form of $\hat{R}$ remains a challenge. We note that the average energy of the quantum particle can be derived as $\langle E\rangle = q\int_\infty (\varphi\rho - \vec{A}\cdot\vec{J})dV + (\hbar^2/2m)\int_\infty \nabla\psi^* \cdot \nabla\psi dV - (q^2/2m)\int_\infty (\vec{A}\cdot\vec{A})\rho dV$. Hence its derivative can also be expressed as

$$d\langle E\rangle/dt = q\int_\infty [(\partial\varphi/\partial t)\rho - (\partial\vec{A}/\partial t)\cdot\vec{J}]dV + q\int_\infty [(\partial\rho/\partial t)\varphi - (\partial\vec{J}/\partial t)\cdot\vec{A}]dV$$
$$+ (d/dt)[(\hbar^2/2m)\int_\infty \nabla\psi^*\cdot\nabla\psi dV - (q^2/2m)\int_\infty (\vec{A}\cdot\vec{A})\rho dV]. \quad (23)$$

By a comparison of Eq. (23) and Eq. (11), one obtains

$$(1/i\hbar)\int_\infty \psi^*[\hat{H},\hat{R}]\psi dV = q\int_\infty [(\partial\rho/\partial t)\varphi - (\partial\vec{J}/\partial t)\cdot\vec{A}]dV$$
$$+ (d/dt)[(\hbar^2/2m)\int_\infty \nabla\psi^*\cdot\nabla\psi dV - (q^2/2m)\int_\infty (\vec{A}\cdot\vec{A})\rho dV]. \quad (24)$$

Equation (24) is another condition for $\hat{R}$, indicating the role of $\partial\rho/\partial t$ and $\partial\vec{J}/\partial t$.

Finally we conclude the work with several remarks. According to Eq. (9) the nonlinear term $\hat{R}\psi = \beta\rho\psi = \beta|\psi|^2\psi$ with $\beta$ being a constant leads to $\langle\vec{F}_r\rangle = 0$, nor does it satisfy many other conditions for $\hat{R}$. Hence the most popular TDNLSE cannot describe the radiation effect, although it has many important applications [13-16]. Many kinds of TDNLSE have been investigated [17-25]. However, the nonlinearity has not been related to the damping effect of the radiation. One characteristic condition for TDNLSE (3) is that the derivatives $\partial\psi/\partial t$ and $\partial\psi^*/\partial t$ should be included in the nonlinear term. Due to this condition, the nonlinearity in TDNLSE (3) is not included in the TDNLSE investigated in Ref. 26, where the nonlinear term is a function of $\psi$ and $\psi^*$. Hence the nonlinearity in TDNLSE (3) is not excluded by the related experiments [26,27]. By including the radiation energy, the energy variation is explicitly presented by Eq. (14), and the problem caused by the absence of the radiation energy is avoided [28]. The small integral on the right of Eq. (22) gives quantitatively the line-broadening of the emission spectrum in both the spontaneous transition and the stimulated transition. Because for the field with a large frequency this integral is very small, the broadening in the stimulated transition seems much smaller than in the spontaneous transition. In conclusion, the nonlinear term in TDNLSE (3) is a physical reality because it originates from the damping effect of the radiation. Although the effect is small, it is decisive for the realization of the stationary states. The TDNLSE hopefully presents the wave-function reduction and transitions between stationary states, at least approximately, without resorting to more exact quantum electrodynamics.